\documentclass{article}

\usepackage{graphicx}
\usepackage{bm}
\usepackage[scaled]{helvet}
\usepackage[margin=1in]{geometry}
\usepackage[nodisplayskipstretch]{setspace}
\usepackage[utf8]{inputenc}
\usepackage{amsmath}
\usepackage{amsfonts}
\usepackage{multicol, booktabs, bbm, color, xcolor}
\usepackage[super]{natbib}
\usepackage{graphicx}
\usepackage{subcaption}
\usepackage{float}
\usepackage[colorlinks=true,linkcolor=black,citecolor=black,urlcolor=black]{hyperref}
\usepackage{colortbl}
\usepackage{xcolor}
\usepackage{etoolbox}
\usepackage{authblk}
\usepackage{gensymb}
\usepackage{titlesec}
\usepackage{lineno}

\graphicspath{{./img/}}
\begin{document}

\setstretch{1}
\title{Excess risk of heat-related hospitalization associated with temperature and PM$_{2.5}$ among older adults}
\date{}
\author[a]{Lauren Mock}
\author[a]{Rachel C. Nethery} 
\author[b]{Poonam Gandhi} 
\author[b]{Ashwaghosha Parthasarathi}
\author[b]{Melanie Rua}
\author[c]{David Robinson}
\author[b,d]{Soko Setoguchi} 
\author[e]{Kevin Josey\thanks{Corresponding author. \\ 
Address: Department of Biostatistics and Informatics, Colorado School of Public Health, 13001 East 17th Place, 3rd Floor, Aurora, CO 80045. E-mail: \texttt{kevin.josey@cuanschutz.edu}
}} 

\affil[a]{Department of Biostatistics, Harvard T.H. Chan School of Public Health, Boston, MA, USA}
\affil[b]{Rutgers University Institute for Health, Healthcare Policy, and Aging Research, New Brunswick, NJ, USA}
\affil[c]{Department of Geography, Rutgers University, Piscataway, NJ, USA}
\affil[d]{Department of Medicine, Rutgers Robert Wood Johnson Medical School, New Brunswick, NJ, USA}
\affil[e]{Department of Biostatistics and Informatics, Colorado School of Public Health, Aurora, CO, USA}

\maketitle

\newpage

\setstretch{2}

\vspace{1cm}

\begin{abstract}
    
    \noindent\textbf{Background:} With rising temperatures and an aging population, understanding how to prevent heat-related illness among older adults will be increasingly crucial. Despite biological plausibility, no study to date has investigated how exposure to fine particulate matter air pollution (PM$_{2.5}$) may contribute to the risk of hospitalization with a diagnosis code indicating heat-related illness, referred to as heat-related hospitalization. This study aims to fill this gap by investigating the independent and combined effects of temperature and PM$_{2.5}$ exposures on heat-related hospitalization risk.

    \noindent\textbf{Methods:} We identified Medicare fee-for-service beneficiaries in the contiguous United States who experienced a heat-related hospitalization between 2008 and 2016. Using a case-crossover design and fitting Bayesian conditional logistic regression models, we characterized the associations of temperature and PM$_{2.5}$ exposures with heat-related hospitalization. We then estimated the relative excess risk due to interaction to quantify the additive-scale interaction of simultaneous exposure to heat and PM$_{2.5}$.

    \noindent\textbf{Results:} We observed 112,969 heat-related hospitalizations among 29,345,820 Medicare beneficiaries in the study sample. Fixing PM$_{2.5}$ at the case day median, the odds ratio for increasing temperature from its case day median to the 95th percentile was 1.05 (95\% CI: 1.03, 1.06). Fixing temperature at the case day median, the odds ratio for increasing PM$_{2.5}$ from its median to the 95th percentile was 1.01 (95\% CI: 0.99, 1.04). The risk due to interaction for simultaneous median-to-95th percentile increases in temperature and PM$_{2.5}$ was 0.03 (95\% CI: 0.01, 0.06).
        
    \noindent\textbf{Conclusions:} Our study is the first to observe synergism between temperature and PM$_{2.5}$ exposures associated with the risk of heat-related hospitalization. These findings highlight the importance of considering air pollution in effective public health and clinical interventions to prevent heat-related illness.
    
\end{abstract}

\newpage

\section{Introduction}

Heat is a leading cause of weather-related illness and death, and its impact is expected to intensify with climate change. \citep{ebi2021hot} Previous research suggests that each year, thousands of deaths in the United States are directly or indirectly attributable to heat exposure. \citep{weinberger2020estimating} Extreme heat events have been associated with increased hospitalizations for conditions that may be triggered or exacerbated by heat, such as respiratory, \citep{anderson2013heat} renal, \citep{fletcher2012association} and cardiovascular \citep{sun2018effects} diseases, as well as conditions that occur as a direct result of heat exposure, \citep{layton2020heatwaves} which are broadly referred to as heat-related illness. Heat-related illness ranges in severity from mild conditions such as heat cramps and syncope to severe, life-threatening conditions like heat stroke. Many of these outcomes are preventable through proactive measures such as avoiding direct sun, maintaining hydration, or temporarily relocating to public cooling centers---actions often prompted by heat alerts. \citep{hasan2021effective, flynn2005older} However, despite these preventive efforts, heat-related emergency department visits in the United States reached record levels during the summer of 2023. \citep{vaidyanathan2024heat}

Older adults are particularly vulnerable to heat due to impaired thermoregulation, social isolation, a higher prevalence of comorbidities, and the use of medications that increase heat sensitivity. \citep{layton2020heatwaves, flynn2005older, khalaj2010health} As the US population ages and extreme heat events become more frequent and intense, \citep{pachauri2014climate} it will become increasingly essential to understand and implement strategies to protect older Americans from heat-related illness.

Previous studies have identified synergistic effects between acute exposure to heat and particulate matter (PM) air pollution on mortality \citep{stafoggia2008does, li2017modification, kioumourtzoglou2016pm2} and on certain cause-specific hospitalizations. \citep{yitshak2018association} These investigations of synergism are especially critical since daily temperature and PM$_{2.5}$ levels are often correlated across much of the United States, even after accounting for seasonal and long-term trends, leading to frequent simultaneous high exposures. \citep{tai2010correlations} Despite biological plausibility, no study has yet investigated the role of PM in heat-related hospitalizations (HRH). PM$_{2.5}$ is small enough to penetrate the lungs and bloodstream, affecting several organ systems. Evidence from epidemiological and toxicological studies shows that acute exposure to PM$_{2.5}$ can impair vascular function, trigger oxidative stress and inflammation, and increase blood pressure. \citep{pryor2022physiological, basith2022impact} These physiological responses may hinder the body's ability to thermoregulate in hot conditions, potentially increasing the risk of heat-related illness when PM$_{2.5}$ levels are high. Understanding the impact of PM$_{2.5}$ on heat-related illness is critical for improving heat alert systems and guiding preparedness efforts.

To address this gap, we investigate the independent and combined effects of temperature and PM$_{2.5}$ on HRHs, defined as hospitalizations due to heat-related illness, among older Medicare fee-for-service (FFS) beneficiaries. Specifically, our analysis focuses on inpatient hospitalizations for heat-related illness during the warm season (June--September) from 2008 to 2016. We employ a case-crossover design and a Bayesian analytic approach to: 1) reduce bias from unmeasured, time-invariant confounders, \citep{basu2005temperature, maclure2000should} and 2) provide interpretable inferences using posterior distributions generated by Markov chain Monte Carlo sampling. We model nonlinear associations between the exposures and outcomes, and we assess synergy between heat and PM$_{2.5}$ by estimating the relative excess risk due to interaction (RERI), an additive-scale interaction quantity. 

\section{Methods}

\subsection{Study population and case definition}

We assessed records from a 50\% random sample of Medicare FFS beneficiaries aged 65 and older who resided in the contiguous United States and were continuously enrolled in Parts A, B, and D for at least 12 months between 2008 and 2016. These records include data on beneficiaries' sex, age, race/ethnicity, Medicaid eligibility status, ZIP code of residence, inpatient admission and discharge dates, and date of death, if applicable. We identified inpatient hospitalizations with a primary or secondary diagnosis of heat-related illness, based on International Classification of Diseases (9th and 10th revisions) diagnosis codes; detailed International Classification of Diseases-9 and -10 codes are provided in Table S1 of Supplementary Material. In the United States, the majority of unscheduled inpatient hospitalizations follow an emergency room visit, although some admissions originate with physician referrals from outpatient settings. \citep{kocher2013changes} We refer to the composite of these indications as HRH. Our analysis is limited to individuals who experienced an HRH during the warm season (June--September), and we included only each beneficiary's first HRH during the study period, excluding any subsequent events.

\subsection{Exposure assessment}

Daily maximum temperature data were obtained from the Parameter-elevation Relationships on Independent Slopes Model, \citep{prism2024prism, daly2008physiographically} which estimates temperature across the contiguous United States using surface station data and geographic factors such as elevation, coastal proximity, and atmospheric conditions. Daily temperature estimates are available at a 4-km resolution. For each ZIP code, we assigned the temperature estimate from the Parameter-elevation Relationships on Independent Slopes Model grid cell whose centroid was nearest to the ZIP code's population-weighted centroid. Daily PM$_{2.5}$ concentrations were derived from a validated ensemble model that integrates ground monitoring data, chemical transport models, satellite observations, and land-use variables. \citep{di2019ensemble, di2025pm25} These estimates, available at a 1-km resolution, were aggregated to the ZIP code level by averaging values from grid cells whose centroids fell within each ZIP code. 

For each individual and date, exposure was defined as: 1) the daily maximum temperature in the individual’s residential 
ZIP code on that date, and 2) the three-day average PM$_{2.5}$ concentration in the individual’s residential 
ZIP code, calculated as the mean concentration from the case day and the two preceding days. We used an extended exposure window for PM$_{2.5}$ to account for the delayed health effects that might be incurred by air pollution exposures, which has been observed previously. \citep{wang2018associations, fernandez2018association} As some previous studies have identified lagged effects of heat on hospitalizations, \citep{delaney2025extreme, gronlund2016vulnerability} we conducted a sensitivity analysis using three-day exposure windows for both temperature and PM$_{2.5}$.

\subsection{Statistical analysis}

We used a case-crossover design to evaluate the effects of acute heat and PM$_{2.5}$ exposures among individuals who experienced an HRH, with each individual serving as their own control. The case day was defined as the admission date of an individual's first HRH during the study period. Control days were assigned bidirectionally within the same calendar month and matched to the same day of the week as the case day, with up to four matched control days per case selected before or after the hospitalization date. To eliminate potentially undue influence from extreme PM$_{2.5}$ concentration outliers, we calculated the 95th percentile of three-day PM$_{2.5}$ exposures across all case and control days and trimmed, or excluded, observations exceeding this threshold. \citep{di2017association, nethery2023air} We also conducted a sensitivity analysis using the 99th percentile as the threshold for trimming.

We implemented the Bayesian analogue to conditional logistic regression using the \texttt{brms} package \citep{burkner2017r} to estimate the associations of temperature and PM$_{2.5}$ with HRH. The primary model comprised three terms: a natural cubic spline for temperature, a natural cubic spline for PM$_{2.5}$, and a linear interaction term. As a sensitivity analysis, we also applied a tensor product spline approach to enable a more complex, nonlinear interaction between the exposures. Detailed descriptions of both approaches are provided in the Supplementary Material. 

We estimated the odds of HRH under various temperature and PM$_{2.5}$ scenarios, reporting posterior means and 95\% Bayesian credible intervals (CIs) derived from the posterior distribution samples of the Markov chain Monte Carlo. While the measures of association from conditional logistic regression models are typically evaluated on a multiplicative scale, multiplicative interactions can be challenging to interpret in a public health context. \citep{rothman2012epidemiology} Instead, we assess the interaction between heat and PM$_{2.5}$ on an additive scale by estimating the relative excess risk due to interaction (RERI). \citep{knol2007estimating, vanderweele2014tutorial} The RERI quantifies the increased risk associated with simultaneous exposure to heat and PM$_{2.5}$ beyond the sum of their individual effects. Because HRH is a rare outcome, the odds ratio-based RERI estimate should closely approximate the conventional RERI, which is a function of relative risks. \citep{vanderweele2014tutorial}

To estimate the RERI for two continuous exposures, we select the median and 95th percentile values from the distributions of temperature and three-day average PM$_{2.5}$ concentrations on case days as contrast levels. This approach to assessing synergism between exposures has been applied in similar epidemiological studies. \citep{fayyad2024air, josey2023retrospective} Additional methodological details are provided in the Supplementary Material. Finally, we also report the interactions between temperature and PM$_{2.5}$ on a multiplicative scale as recommended by VanderWeele and Knol \citep{vanderweele2014tutorial} as well as  Vandenbroucke et al, \citep{vandenbroucke2007strengthening} estimating this measure using posterior means and 95\% Bayesian CIs from the linear interaction coefficient.

All analyses were completed in R (version 4.4.0; Vienna, Austria); \citep{r2021r} code to reproduce these analyses is available at \url{https://github.com/laurenmock/temp_pm25_heat-hosp}.

Our study aims and data stewardship plan were approved by the Rutgers University Institutional Review Board (Pro20170001685). Medicare patient data are protected by our data usage agreement with the Centers for Medicare and Medicaid, and were stored and analyzed with minimal risk of breach of data confidentiality. Use of Medicare data does not require informed consent from individual beneficiaries.

\section{Results}

\subsection{Main analysis}

Among 29,345,820 Medicare beneficiaries in our sample, we observed at least one HRH among 112,969 unique individuals. Each individual’s first HRH served as a case in our case-crossover study. Heat-related illness was the primary diagnosis 
for 60,053 hospitalizations (53.2\%) and the secondary diagnosis for the remaining hospitalizations. The mean length of 
stay in the hospital was 3.5 (standard deviation: 4.5) days. The characteristics of the individuals who experienced an HRH are detailed in Table \ref{tab:table1}. Most of the hospitalized individuals were women (68.2\%) and white (82.3\%), with 22.9\% eligible for Medicaid. The mean age at the time of hospitalization was 81.7 years. On case days and control days, the mean maximum daily temperatures were 29.2$\degree$C and 29.1$\degree$C, respectively. Before trimming, the case day distribution of three-day PM$_{2.5}$ concentrations was heavily right skewed, with exposures as high as 120.4 $\mu g/m^3$. Case and control days with three-day PM$_{2.5}$ exposures above 18.4 $\mu g/m^3$, the 95th percentile of the distribution, were trimmed; this resulted in a mean PM$_{2.5}$ concentration of 9.4 $\mu g/m^3$ on case days and 9.3 $\mu g/m^3$ on control days. The full distributions of temperature and PM$_{2.5}$ case day exposures are displayed in Figures S1 and S2 in the Supplementary Material.

% latex table generated in R 4.5.0 by xtable 1.8-4 package
% Fri Oct 31 11:10:50 2025
\begin{table}[ht]
\centering
\begin{tabular}{ll}
  \hline
  & N (\%) \\ 
  \hline
Total & 112,969 \\ 
  Sex &  \\ 
  \hspace{10pt}Male & 35,930 (31.8\%) \\ 
  \hspace{10pt}Female & 77,039 (68.2\%) \\ 
  Age category &  \\ 
  \hspace{10pt}65-74 & 19,165 (17.0\%) \\ 
  \hspace{10pt}75-84 & 45,868 (40.6\%) \\ 
  \hspace{10pt}85+ & 47,936 (42.4\%) \\ 
  Race/ethnicity &  \\ 
  \hspace{10pt}White & 92,975 (82.3\%) \\ 
  \hspace{10pt}Black/African-American & 13,123 (11.6\%) \\ 
    \hspace{10pt}Hispanic & 3,178 (2.8\%) \\ 
  \hspace{10pt}Asian/Pacific Islander & 2,006 (1.8\%) \\ 
  \hspace{10pt}American Indian/Alaska Native & 650 (0.6\%) \\ 
  \hspace{10pt}Other & 1,037 (0.9\%) \\ 
  Medicaid eligibility &  \\ 
  \hspace{10pt}Ineligible & 87,059 (77.1\%) \\ 
  \hspace{10pt}Eligible & 25,910 (22.9\%) \\[6pt]
  \hline
   & Mean (SD) \\
  \hline
  Daily maximum temperature ($\degree$C) &  \\
  \hspace{10pt}Case days & 29.2 (5.0) \\
  \hspace{10pt}Control days & 29.1 (5.1) \\ 
  Three-day PM$_{2.5}$ ($\mu g/m^3$) &  \\
  \hspace{10pt}Case days & 9.4 (3.7) \\ 
  \hspace{10pt}Control days & 9.3 (3.6) \\ 
   \hline
\end{tabular}
\caption{Characteristics of individuals included in the analysis and exposures on case and control days. Case day temperature exposure is defined as the maximum daily temperature on the day of hospitalization. Case day PM\(_{2.5}\) exposure is defined as the three-day mean PM\(_{2.5}\) exposure on the day of hospitalization and the two preceding days.\newline
SD, standard deviation.} 
\label{tab:table1}
\end{table}

\begin{figure}[H]
    \centering
    \includegraphics[width=16cm]{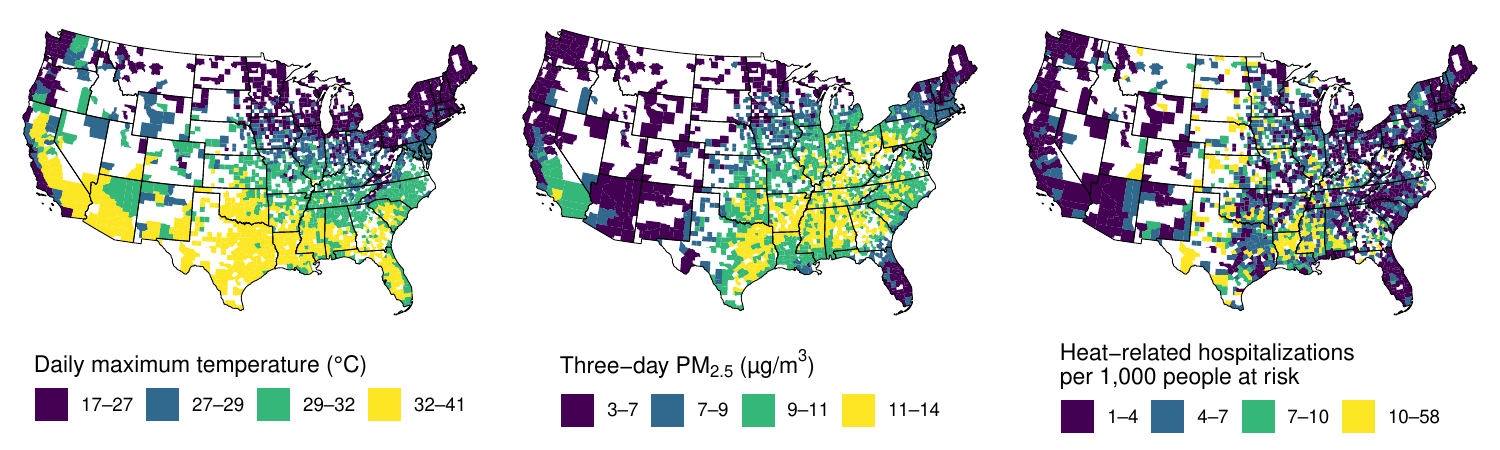}
    \caption{From left to right, aggregated county-level maps of 1) mean case day maximum temperature, 2) mean case day three-day PM$_{2.5}$ concentrations, and 3) number of heat-related hospitalizations per 1000 Medicare FFS beneficiaries at risk of hospitalization during the study period. Counties with 10 or fewer heat-related hospitalizations observed during the study period are suppressed for confidentiality and are displayed in white.}
    \label{fig:temp_pm25_hosp_maps}
\end{figure}

Temperature exposure on case days was highest in the South and Southwest regions of the United States (Figure \ref{fig:temp_pm25_hosp_maps}, left panel), while PM$_{2.5}$ exposure concentrations were highest in the South, Midwest, and Southern California (Figure \ref{fig:temp_pm25_hosp_maps}, middle panel). The rates of HRH per 1000 Medicare FFS beneficiaries at risk varied substantially across the United States, with the highest rates in the Central United States (Figure \ref{fig:temp_pm25_hosp_maps}, right panel).

\begin{figure}[h]
    \centering
    \includegraphics[width=16cm]{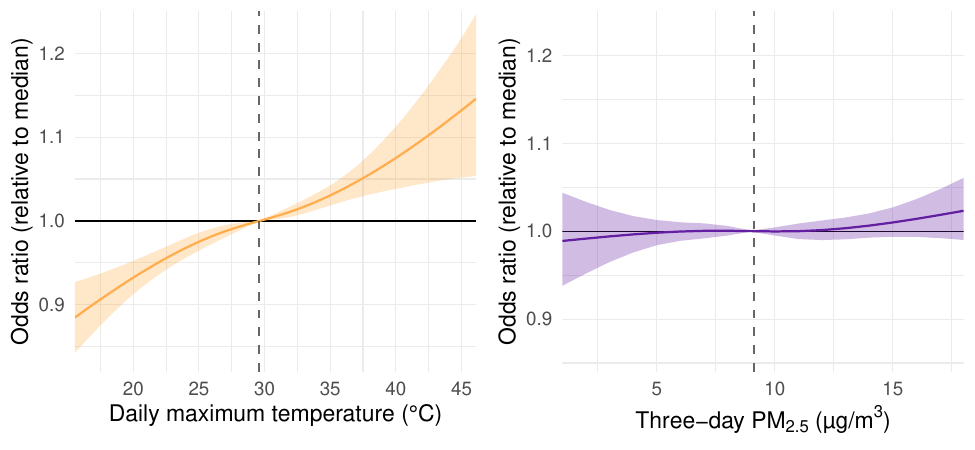}
    \caption{Independent nonlinear effects of temperature and PM$_{2.5}$ on heat-related hospitalization. The left panel displays the odds ratio of heat-related hospitalization comparing the median case day temperature (29.6$\degree$C), shown with a vertical dashed line, to various alternative temperatures across the \textit{x} axis, while holding PM$_{2.5}$ exposure fixed at the median (8.9 $\mu g/m^3$). The right panel displays the odds ratio of heat-related hospitalization comparing the median case day PM$_{2.5}$ exposure, shown with a vertical dashed line, to various alternative concentration levels across the \textit{x} axis, while holding temperature exposure fixed at the median.}
    \label{fig:independent_effects_main}
\end{figure}

Figure \ref{fig:independent_effects_main} shows the independent effects of heat and PM$_{2.5}$ on HRH, with PM$_{2.5}$ held at its case day median to examine the effect of heat, and temperature held at its case day median to examine the effect of PM$_{2.5}$. Maximum daily temperature exhibited a strong positive association with odds of HRH (Figure \ref{fig:independent_effects_main}, left panel). Specifically, when PM$_{2.5}$ concentrations are fixed at 8.9 $\mu g/m^3$, increasing the temperature from 29.6$\degree$C (median) to 36.9$\degree$C (95th percentile) yields an odds ratio of 1.05 (95\% CI: 1.03, 1.06) (Figure \ref{fig:or_reri} and Table S2 in the Supplementary Material). Although a positive association was observed between three-day PM$_{2.5}$ and HRH, credible intervals were wide and covered one across the range of PM$_{2.5}$ levels (Figure \ref{fig:independent_effects_main}, right panel). When temperature was fixed at 29.6$\degree$C, an increase in PM$_{2.5}$ concentrations from 8.9 $\mu g/m^3$ (median) to 16.1 $\mu g/m^3$ (95th percentile) yielded an odds ratio of 1.01 (95\% CI: 0.99, 1.04) (Figure \ref{fig:or_reri} and Table S2 in the Supplementary Material).

Furthermore, we observed an additive interaction between temperature and PM$_{2.5}$. A simultaneous increase in both temperature and PM$_{2.5}$ from their median values to their respective 95th percentile quantities yielded an odds ratio of 1.09 (95\% CI: 1.06, 1.12) (Figure \ref{fig:or_reri} and Table S2 in the Supplementary Material). Using these odds ratios to approximate the RERI, we found that the approximate risk increase was 0.03 (95\% CI: 0.01, 0.06) (Figure \ref{fig:or_reri} and Table S2 in the Supplementary Material), indicating synergism between the two exposures resulting in excess HRHs. We visualized synergism between these two exposures by comparing the odds of HRH at median exposure levels to the odds of HRH across a grid of exposure values (Figure \ref{fig:risk_surface_main}). Temperature was strongly associated with HRH, but on hot days, the odds of HRH also increased with higher concentrations of PM$_{2.5}$. On cooler days, however, we see a small protective effect of PM$_{2.5}$. Since temperature and PM$_{2.5}$ were often correlated (as shown in Figure S2), we have limited data to assess the effect of elevated PM$_{2.5}$ at lower temperatures during the warm season.

\begin{figure}[H]
    \centering
    \includegraphics[width=16cm]{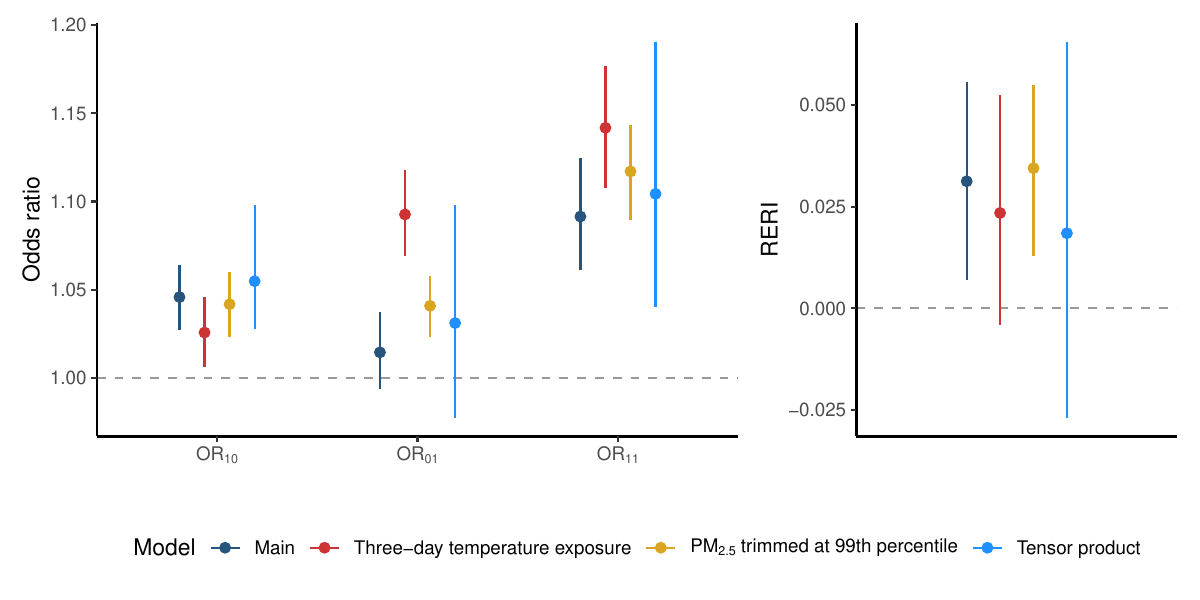}
    \caption{Comparison of the main and sensitivity analyses. The left panel displays odds ratios with 95\% Bayesian credible intervals associated with (1) increasing temperature from the median to the 95th percentile while holding PM$_{2.5}$ at the median, (2) increasing PM$_{2.5}$ from the median to the 95th percentile while holding temperature at the median, and (3) increasing both temperature and PM$_{2.5}$ from the median to the 95th percentile simultaneously. The right panel displays the relative excess risk due to interaction (RERI) with 95\% Bayesian credible intervals. The same exposure contrasts were used across models. All numerical results are presented in Table S2 in the Supplementary Material.}
    \label{fig:or_reri}
\end{figure}

Although we were primarily interested in assessing the additive interaction with the RERI, we also estimated the interaction on a multiplicative scale. The estimated odds ratio for the linear interaction term was 1.0003 (95\% CI: 1.0001, 1.0006). Both the additive and multiplicative interaction estimates were positive, with credible intervals that did not cover the null value.

\begin{figure}[H]
    \centering
    \includegraphics[width=10cm]{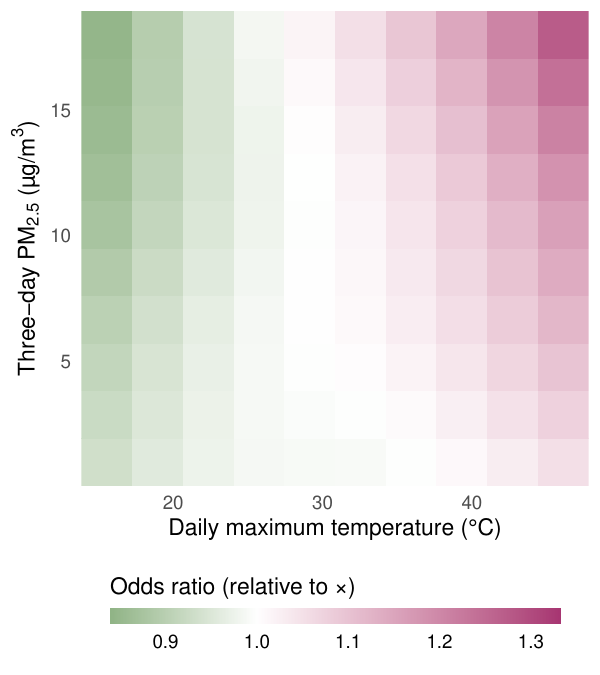}
    \caption{Synergistic effects of temperature and PM$_{2.5}$ across a range of exposure values. The tile color indicates the odds ratio of heat-related hospitalization for a given pair of temperature and PM$_{2.5}$ exposures versus the median temperature and PM$_{2.5}$ exposures (marked on the grid with an $\times$).}
    \label{fig:risk_surface_main}
\end{figure}

\subsection{Sensitivity analyses}

To investigate the interaction between temperature and PM$_{2.5}$ further and assess the robustness of our findings, we fit three sensitivity analyses. Note that we used the median and 95th percentile contrasts defined in the main analysis across all sensitivity analyses to allow for direct comparisons across models.

First, we extended the temperature window from one day to three days and observed attenuated estimated associations between temperature and HRH with wider credible intervals relative to the main analysis (Figure 3 and Figures S3 and S4 in the Supplementary Material). Second, we modified the trimming threshold from the 95th percentile of the three-day PM$_{2.5}$ distribution (18.4 $\mu g/m^3$) to the 99th percentile (23.7 $\mu g/m^3$). The findings indicate a much stronger positive association between PM$_{2.5}$ and HRH than that seen in the main analysis, with narrow credible intervals that do not include one (Figure 3 and Figure S3 in the Supplementary Material). Finally, we developed an additional model using a tensor product to allow for a more flexible, nonlinear interaction between these exposures. This approach resulted in more linear estimates of the independent effects of temperature and PM$_{2.5}$ compared with the other analyses (Figure S3 in the Supplementary Material). The estimated RERIs from all three analyses were positive, but the credible intervals from the sensitivity analyses with a longer temperature exposure and a tensor product included the null value of zero (Figure 3).

\section{Discussion}

\subsection{Summary and key findings}

To our knowledge, this is the first study to investigate the combined (and synergistic) effects of acute exposure to high temperatures and elevated PM$_{2.5}$ on HRHs. Comparing median and 95th percentile exposure levels, we found a positive association between temperature and HRH. The association between PM$_{2.5}$ and HRH was also positive, although the credible interval covered the odds ratio scale null value of one. Finally, we identified evidence of an additive interaction between temperature and PM$_{2.5}$. 

Three sensitivity analyses demonstrate robustness in our findings and provide additional insights. Most notably, the analysis with a longer temperature exposure window indicated a weaker relationship between temperature and HRH, suggesting that the effects of temperature on HRH are more immediate. Additionally, the wider credible intervals from the tensor model reflect increased uncertainty in the estimates when allowing for greater flexibility in the interaction structure.

Our findings offer the first empirical, epidemiological evidence that PM$_{2.5}$ may increase the risk of HRH. In contrast to prior work, which has largely framed temperature as a modifier of the health effects of air pollution, \citep{li2017modification, areal2022effect, anenberg2020synergistic} our focus on explicit HRHs reveals that PM$_{2.5}$ may also modify the health effects of heat.

\subsection{Impact}

Heat is the deadliest form of extreme weather in the United States, \citep{national2024weather} and rates of heat-related illness continue to climb \citep{vaidyanathan2024heat, osborne2023trends} even though these illnesses are often preventable. \citep{sorensen2022treatment, gauer2019heat} Understanding how to protect vulnerable individuals, including older adults, from extreme heat events is a crucial component of climate change adaptation and an urgent public health priority. \citep{vaidyanathan2024heat}

The National Weather Service issues localized heat alerts to initiate individual- and community-level preparations on unusually warm days. \citep{national2024heat} Our findings indicate that additional caution is necessary on hot days when PM$_{2.5}$ levels are high, and it may be useful to consider PM$_{2.5}$ when issuing heat alerts. However, the efficacy of heat alerts alone in minimizing illness and death is uncertain. \citep{weinberger2018effectiveness, toloo2013evaluating} Previous research highlights that high-risk individuals must recognize their own vulnerabilities to heat, possess the knowledge and resources necessary to protect themselves, and take action to stay safe. \citep{toloo2013evaluating} Interestingly, previous studies report that many older adults recognize age as a risk factor for heat-related illness but do not recognize high temperatures as a threat to their personal health. \citep{widerynski2017use} Clinicians could play an important role in helping these individuals understand their risk. \citep{sorensen2022treatment}

Our results also emphasize the importance of minimizing personal exposure to particulate matter, particularly on unusually warm days. For example, open windows may allow more ambient particulate matter into homes \citep{lin2013reducing} and further increase the risk of heat-related illness. These findings also contribute to the growing body of evidence that PM$_{2.5}$ is linked to a range of negative health effects, underscoring the importance of reducing ambient PM$_{2.5}$, particularly with a warming planet.

\subsection{Epidemiological evidence and biological plausibility of 
synergism}

This study supports mounting epidemiological evidence of synergism between temperature and PM$_{2.5}$. Several studies have assessed effects on mortality; for example, a case-crossover study from Italy \cite{stafoggia2008does} identified positive but generally nonstatistically significant interactions between temperature and PM$_{10}$, which includes both PM$_{2.5}$ and larger particles, on mortality, and a meta-analysis by Li and colleagues \cite{li2017modification} also reported that extreme temperatures modify the effect of PM$_{10}$ on nonaccidental and cardiovascular mortality. An examination of PM$_{2.5}$ and mortality in US cities \cite{kioumourtzoglou2016pm2} revealed a stronger association between long-term PM$_{2.5}$ exposure and mortality in warmer cities. A smaller number have assessed hospitalization outcomes, including a recent study in New England that identified synergism between short-term exposures to temperature and PM$_{2.5}$ for both respiratory and cardiac hospital admissions. \citep{yitshak2018association} Our study contributes to this growing body of work and extends scientific understanding of synergism between these short-term exposures to HRHs, providing additional insight into the pathways between simultaneous temperature and air pollution exposures and morbidity and mortality.

Furthermore, it is biologically plausible that PM$_{2.5}$ may interfere with the body’s ability to respond to high temperatures. Simultaneous exposure to heat and PM$_{2.5}$ can exacerbate heat-related health outcomes through shared pathophysiological mechanisms, including autonomic dysfunction and heightened inflammatory responses. Both heat and PM$_{2.5}$ cause autonomic dysfunction by increasing sympathetic drive and suppressing parasympathetic activity. \citep{taylor2020air, yamamoto2007evaluation} This imbalance leads to a 4\%–7\% reduction in heart rate variability and an increase in heart rate by 5–9 beats per minute. \citep{shah2019heart, florea2014autonomic, jensen2015relationship} In addition, blood pressure increases due to autonomic vasoconstriction from PM$_{2.5}$ exposure, with each 10 $\mu g/m^3$ increment correlating with a 9-mm Hg rise in systolic pressure. \citep{dvonch2009acute} Furthermore, increased sweating via sympathetic activation in response to heat elevates blood pressure by increasing plasma viscosity by 2\% per $\degree$C. \citep{sakurai2004plasma} Thus, the combined cardiovascular stress disrupts hemodynamic regulation, compromising the body's thermoregulatory function and exacerbating heat stress.

Heat and PM$_{2.5}$ can also increase systemic inflammatory responses by forming reactive oxygen species within the cells. The resulting oxidative stress activates redox-sensitive transcription factors, driving the production and release of pro-inflammatory mediators such as tumor necrosis factor-alpha, interleukin-6, and interleukin-1$\beta$. \citep{ebi2021hot, sangkham2024update} A 10 $\mu g/m^3$ increase in PM$_{2.5}$ exposure corresponds to a five-fold rise in these cytokines, and a 2$\degree$C elevation in ambient temperature can boost their expression by up to four-fold. \citep{hu2024effect, park2024mechanism} Elevated cytokines also disrupt endothelial homeostasis and increase vascular permeability. \citep{park2024mechanism, pope2016exposure, briet2007endothelial} The sustained up-regulation of cytokines affects the hypothalamic thermoregulatory center, while impaired vascular permeability decreases the efficacy of peripheral vasodilatory response and increases the risk of heat-related hospital admissions. \citep{vybiral2005pyrogenic}

\subsection{Limitations and strengths}

We acknowledge several limitations to our study. As with any case-crossover study, the results may not fully generalize to other populations or demographic groups; that is, the group of individuals who experienced an HRH is comprised of a small percentage of all Medicare FFS beneficiaries, and results drawn from this sample may not generalize to the overall US population. Furthermore, it is likely that PM$_{2.5}$ partially mediates the relationship between temperature and various health conditions, \citep{tran2023short, ham2025mediation} including heat-related illness, so our estimate of the independent effect of temperature may not fully capture more complex pathways involving PM$_{2.5}$. Consequently, interpreting temperature and PM$_{2.5}$ as entirely independent exposures may be overly simple, as these exposures may influence each other’s effects on health outcomes in complex ways. We attempted to capture this with an interaction term, but this may not sufficiently capture the true relationship between the exposures. Moreover, due to the complexities of quantifying synergism between two continuous, time-varying exposures, our model did not capture daily exposures over time or interactions between lagged exposures that could be incorporated with a distributed lag model.

It is also possible that the observed relationship between temperature and PM$_{2.5}$ is confounded by other pollutants such as ground-level ozone, which may be correlated with PM$_{2.5}$ particularly in warm weather, \citep{zhu2019correlations} although previous studies have produced mixed results. \citep{bell2007potential} Finally, maximum daily temperature does not account for other meteorological variables such as humidity and wind speed, which may influence temperature perception; however, previous research indicates the choice of heat metric does not substantially alter epidemiological results. \citep{spangler2023does}

This study also has several important strengths. We leverage a large and diverse sample of Medicare FFS beneficiaries, enabling robust estimates of risk across a range of environmental conditions. By applying a case-crossover design, the study effectively controls for confounders that are time-invariant or change slowly over time, allowing for more precise inferences of acute exposure effects. Moreover, our use of a flexible statistical model that incorporates both additive and interactive effects of temperature and PM$_{2.5}$ provides valuable insights into the complex, real-world dynamics between these environmental exposures and their impacts on human health. Together, these strengths support the reliability of our findings and their relevance for guiding targeted public health interventions to mitigate the risks of HRH in vulnerable populations.

\subsection{Future work}

As this study is the first to provide epidemiological evidence for synergism between the effects of temperature and PM$_{2.5}$ on heat-related illness, additional work is needed to understand how these exposures interact. It has been well-established that the dangers of heat vary by geographic region and also time of year, \citep{anderson2011heat} since individuals adapt behaviorally and biologically to heat. An analysis that accounts for climate type or region may uncover differences in these effects by location. For example, the threat of PM$_{2.5}$ may be greater in areas where fewer people have access to air conditioning and are thus more likely to be exposed to ambient air pollution on hot days. PM$_{2.5}$ is also a diverse category of pollutants that includes a wide array of chemical components from various sources, and these components vary widely by location. Evidence suggests that certain components are more harmful than others. \citep{hao2023national, masselot2022differential, feng2024long} Finally, since this study focuses on older adults, there is a need to assess potential synergism in younger vulnerable individuals, such as children or adults who work outdoors. This study is a first step in understanding the complex synergism between the effects of temperature and PM$_{2.5}$ on heat-related illness, and additional research is essential to understand the risk of simultaneous exposure.

% put tables and figures here so they're easy to delete in word doc
% but the numbers will still show up correctly

\newpage

\noindent \textbf{Acknowledgments:} All computations presented here were supported by the Office of Research Computing at the Rutgers University Institute for Health, Health Care Policy and Aging.

\vspace{1cm}

\noindent \textbf{Conflicts of interest:} S.S. received research funding from Pfizer Inc., Pfizer Japan, Bristol Myers Squibb, and Daiichi Sankyo and served as a consultant for Pfizer Japan and Merck \& Co., Inc. Other authors have no potential conflicts of interest to report.

\vspace{1cm}

\noindent \textbf{Data availability:} To protect the privacy of Medicare beneficiaries, our data usage agreement with the Centers for Medicare \& Medicaid Services (CMS) prohibits us from sharing patient data publicly. However, interested parties may purchase these data from CMS through the Research Data Assistance Center (ResDAC). The temperature and PM$_{2.5}$ daily exposure data we used in our analysis are publicly available at \url{https://prism.oregonstate.edu/} and \url{https://doi.org/10.7910/DVN/58C6HG}, respectively. Finally, our R analysis scripts are available at \url{https://github.com/laurenmock/temp_pm25_heat-hosp}.

\vspace{1cm}

\noindent \textbf{Source of funding:} This work was funded by the National Institutes of Health grants R01AG060232 (principal investigator: S.S.) and 1K01ES032458 (principal investigator: R.N.) and an Amazon Web Services grant on AI/ML for Identifying Social Determinants of Health.

\newpage

% Reset numbering for equations, figures, & tables
\renewcommand{\theequation}{S\arabic{equation}}
\renewcommand{\thefigure}{S\arabic{figure}}
\renewcommand{\thetable}{S\arabic{table}}
\setcounter{section}{0}  % reset section counter
\setcounter{equation}{0}  % Reset equation counter
\setcounter{figure}{0} % Reset figure counter
\setcounter{table}{0}  % Reset table counter

\begin{center}
  {\LARGE \bfseries Supplementary materials for \\``Excess risk of heat-related hospitalization associated with temperature and PM$_{2.5}$ among older adults''}
\end{center}
\vspace{1em}

\section{Methods}

\subsection{ICD codes}

Heat-related hospitalizations were identified when the primary or secondary diagnosis code of an inpatient hospitalization indicated heat-related illness. We used the following ICD-9 and ICD-10 codes to define heat-related illness:

\begin{table}[ht]
\centering
\begin{tabular}{lll}
  \hline
  & ICD-9 & ICD-10 \\
  \hline
  Effects of heat and light & 992 & T67 \\
  Exposure to excessive natural heat & E900 (excluding E900.1) & X30 \\
  Dehydration & 2765 & E86 \\
   \hline
\end{tabular}
\caption{Inpatient hospitalization diagnosis codes used to identify heat-related hospitalizations.} 
\label{tab:icd_codes}
\end{table}

\subsection{Main analysis: natural cubic splines}

We fit natural cubic splines for the effects of temperature and PM$_{2.5}$ using the \texttt{ns()} function from the \texttt{splines} R package. Each spline had three degrees of freedom, with knot placement performed in an a priori fashion---the default for the \texttt{ns()} function.

\subsection{Sensitivity analysis: tensor product}

As a sensitivity analysis, we also fit a model using a tensor product smooth to more flexibly capture the interaction between temperature and PM$_{2.5}$. Tensor product smooths combine separate marginal smooth terms for each covariate, allowing each to have its own basis functions and smoothness penalties. Specifically, we implemented a \texttt{t2}-style tensor product smooth from the \texttt{mgcv} R package, \citep{wood2013straightforward} which is compatible with Bayesian modeling in \texttt{brms}. Note that the alternative \texttt{te}-style tensor product smooths, which apply penalties differently, are currently not supported in \texttt{brms}. For further details and guidance on implementing tensor product smooths (\texttt{t2} and \texttt{te}) using \texttt{mgcv}, see the tutorial by Pedersen et al. \citep{pedersen2019hierarchical}

\subsection{OR and RERI estimation details}

In a case-crossover study, we use conditional logistic regression to assess how temperature ($T$) and PM$_{2.5}$ ($A$) exposures affect the odds of heat-related hospitalization. Let $Y_{ij}$ indicate the outcome for observation $j = 1,2,\ldots,m_i$ in beneficiary $i = 1,2,\ldots,n$ (e.g., for each beneficiary, $Y_{ij}=1$ for the case day and $0$ for the control days). We model the log-odds (logit) of the event for a given day’s exposures as:
\begin{equation}
\lambda_i(T_{ij},A_{ij}) = \log\left[\frac{\Pr\left(Y_{ij} = 1 \mid T_{ij},A_{ij},\alpha_i\right)}{\Pr\left(Y_{ij} = 0 \mid T_{ij}, A_{ij}, \alpha_i\right)}\right] = \alpha_i + f(T_{ij}) + g(A_{ij}) + h(T_{ij}, A_{ij}).
\end{equation}
Here $f(\cdot)$ and $g(\cdot)$ are flexible functions (e.g., spline terms) capturing potentially nonlinear effects of temperature and PM$_{2.5}$, and $h(T,A)$ is an interaction term allowing the combined effect of $T$ and $A$ to deviate from additivity on the log-odds scale. In our primary model, we assume a linear interaction $h(T,A) = \gamma T \times A$. In a sensitivity analysis we assume $h(T,A)$ is a tensor product of two spline basis functions. The term $\alpha_i$ is a stratum-specific intercept for each beneficiary, which absorbs baseline risk differences between beneficiaries. Because we condition on each beneficiary, these $\alpha_i$ terms are neither estimated nor reported---in other words, the model does not yield an overall baseline probability of hospitalization. Instead, the regression focuses on estimating the relative effect of exposures within each beneficiary. As a result, the exposure coefficients have a conditional interpretation (i.e., comparing odds of the event between two exposure levels for the same beneficiary) rather than a population-average effect.

The fitted model yields estimates that resemble the log-odds of heat-related hospitalization for any combination of temperature and PM$_{2.5}$ values for a given beneficiary $i = 1,2,\ldots,n$. Despite the similarities, the fitted values are not actual predictions of the log-odds given the design of the case-crossover study. However, we can still estimate the odds ratio for the outcome from these quantities by contrasting the fitted values of two exposure scenarios. Supposing $\alpha_i$ is the same in both contrasts, exponentiating this difference yields an identifiable odds ratio (OR). For example, consider the distributions of observed case day temperature and PM$_{2.5}$ exposures. Let $t_0$ be the median temperature and $t_1$ be a higher temperature (e.g., the 95th percentile), and similarly define $a_0$ and $a_1$ for PM$_{2.5}$. The effect of a higher temperature (vs. median temperature) on hospitalization odds, at a common PM$_{2.5}$ level, can be expressed as: 
\begin{equation}
OR_{10} = \exp\left[\lambda_i(t_1,a_0) - \lambda_i(t_0,a_0) \right].
\end{equation}
Without needing to predict $\alpha_i$, this quantity is identifiable and estimated by 
\begin{equation}
\widehat{OR}_{10} = \exp\left[\hat{f}(t_1) + \hat{h}(t_1,a_0) - \hat{f}(t_0) - \hat{h}(t_0,a_0) \right]
\end{equation}
where $\hat{f}(\cdot)$ and $\hat{h}(\cdot)$ are fit with conditional logistic regression. Similarly, we can examine the odds of a heat-related hospitalization attributable to PM$_{2.5}$ by estimating 
\begin{equation}
OR_{01} = \exp\left[\lambda_i(t_0,a_1) - \lambda_i(t_0,a_0) \right]
\end{equation}
with the plugin estimate 
\begin{equation}
\widehat{OR}_{01} = \exp\left[\hat{g}(a_1) + \hat{h}(t_0,a_1) - \hat{g}(a_0) - \hat{h}(t_0,a_0) \right]. 
\end{equation}
Like $\hat{f}(\cdot)$ and $\hat{h}(\cdot)$, $\hat{g}(\cdot)$ is fit with conditional logistic regression. Finally, we can also examine the joint effect of high temperature and high PM$_{2.5}$ by comparing the odds of an event where both exposures are high to a baseline odds when both exposures are at median levels. This quantity is identified as 
\begin{equation}
OR_{11} = \exp\left[\lambda_i(t_1,a_1) - \lambda_i(t_0,a_0) \right]
\end{equation}
which is estimated with
\begin{equation}
\widehat{OR}_{11} = \exp\left[\hat{f}(t_1) + \hat{g}(a_1) + \hat{h}(t_1,a_1) - \hat{f}(t_0) - \hat{g}(a_0) - \hat{h}(t_0,a_0) \right].
\end{equation}

To quantify departure from additive effects from predictions on a relative risk scale, we can compute the relative excess risk due to interaction (RERI). In our context, we can define exposures as “high” vs “baseline” (e.g. $t_1$ vs $t_0$, and $a_1$ vs $a_0$) and use the odds ratios as approximations for the relative risk predictions. The RERI is then computed using already estimated values with
\begin{equation}
\widehat{RERI} = \widehat{OR}_{11} - \widehat{OR}_{10} - \widehat{OR}_{01} + 1.
\end{equation}

\section{Exposure distributions}

\begin{figure}[H]
    \centering
    \includegraphics[width=16cm]{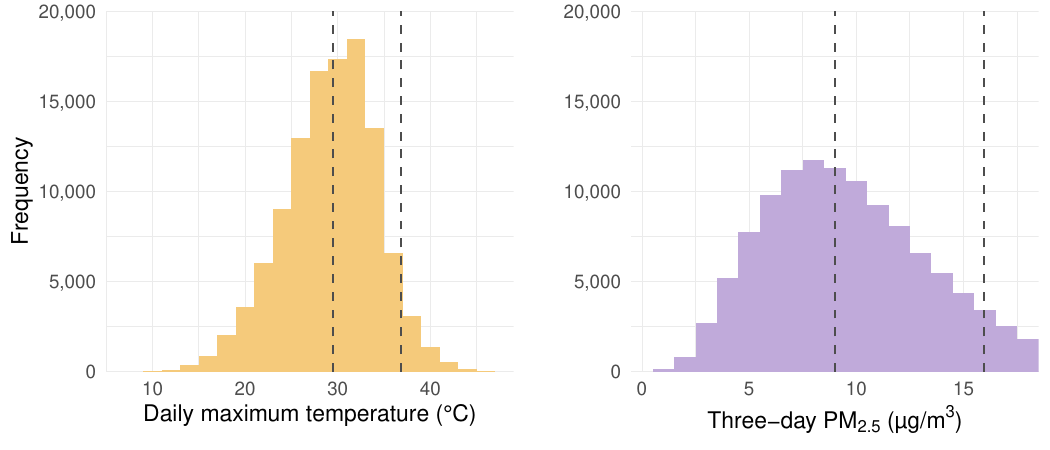}
    \caption{Case day distributions of daily maximum temperature and three-day average PM$_{2.5}$ exposures. The dashed lines indicate the median and 95th percentiles of each exposure, which we use to estimate the RERI. Note that days on which PM$_{2.5}$ exceeded 18.4 $\mu g/m^3$ were excluded from the analysis. Prior to trimming, the maximum case day three-day average PM$_{2.5}$ exposure was 120.4 $\mu g/m^3$.}
    \label{fig:temp_pm25_dist}
\end{figure}

\begin{figure}[H]
    \centering
    \includegraphics[width=12cm]{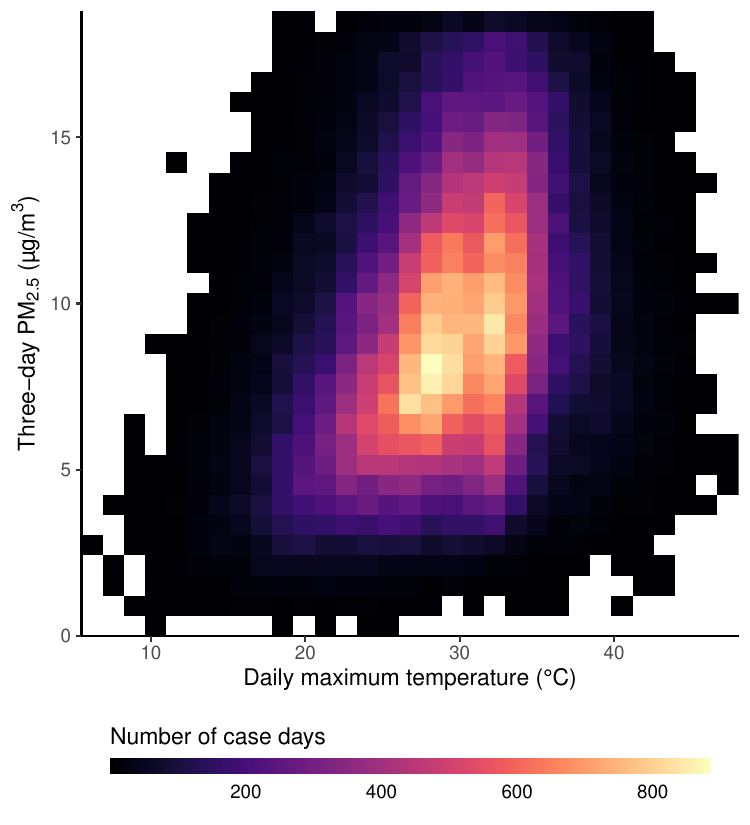}
    \caption{Case day joint daily maximum temperature and three-day average PM$_{2.5}$ exposures. Each case day is assigned to the pixel that covers its temperature and PM$_{2.5}$ exposures; color indicates the number of case days falling within each pixel. White pixels indicate the absence of case days with a given temperature and PM$_{2.5}$ joint exposure.}
    \label{fig:temp_vs_pm25}
\end{figure}

\subsection{Sensitivity analysis results}

We conducted three sensitivity analyses. First, we considered three-day average temperature exposures instead of same-day exposures. Second, we trimmed PM$_{2.5}$ exposures at the 99th percentile instead of the 95th percentile. Lastly, we used a tensor product as opposed to natural cubic splines with an interaction term to capture non-linear interaction between exposures. 

\begin{figure}[H]
    \centering
    \includegraphics[width=16cm]{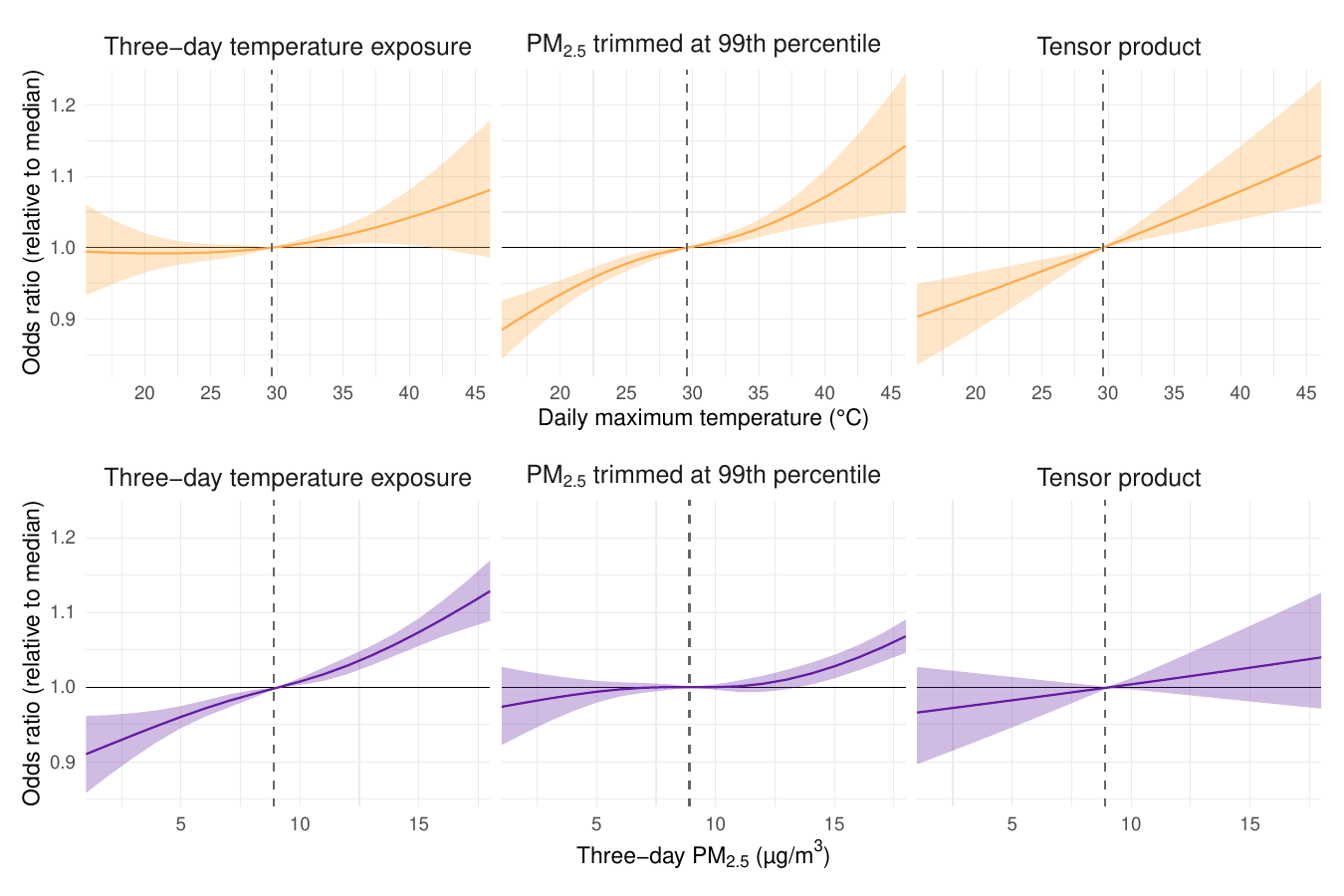}
    \caption{Equivalent to Figure 2 with the independent nonlinear effects of temperature and PM$_{2.5}$ on heat-related hospitalization from the three sensitivity analyses. The top row displays the odds ratio of heat-related hospitalization comparing the median case day temperature (29.6$\degree$C), shown with a vertical dashed line, to various alternative temperatures across the x-axis, while holding PM$_{2.5}$ exposure fixed at the median (8.9 $\mu g/m^3$). The bottom row displays the odds ratio of heat-related hospitalization comparing the median case day PM$_{2.5}$ exposure, shown with a vertical dashed line, to various alternative concentration levels across the x-axis, while holding temperature exposure fixed at the median.}
    \label{fig:sensitivity_independent effects}
\end{figure}

\begin{figure}[H]
    \centering
    \includegraphics[width=16cm]{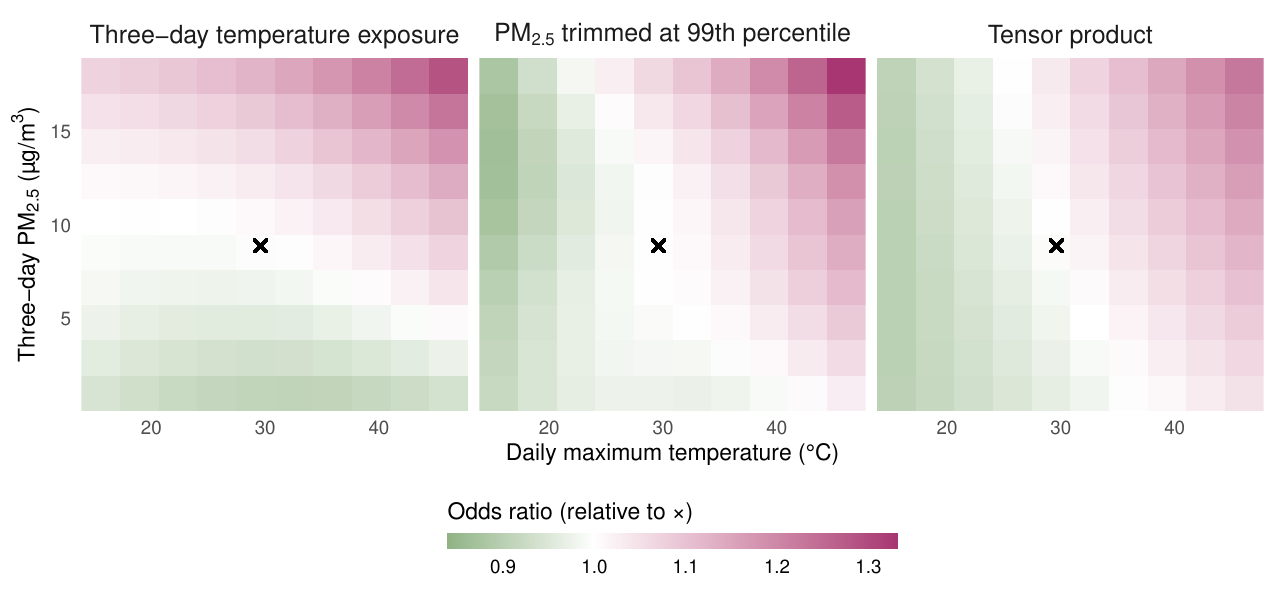}
    \caption{Equivalent to Figure 4 with the synergistic effects of temperature and PM$_{2.5}$ across a range of exposure values from the three sensitivity analyses. The tile color indicates the odds ratio of heat-related hospitalization for a given pair of temperature and PM$_{2.5}$ exposures versus the median temperature and PM$_{2.5}$ exposures from the main analysis (marked on the grid with an $\times$).}
    \label{fig:sensitivity_risk_surface}
\end{figure}

% latex table generated in R 4.5.0 by xtable 1.8-4 package
% Fri Oct 31 14:48:26 2025
\begin{table}[ht]
\centering
\begin{tabular}{lllll}
  \hline
  \\[-2pt]
  \multicolumn{1}{c}{ } & \multicolumn{4}{c}{Estimate (95\% credible interval)} \\[6pt]
  \cline{2-5}
  \\[-2pt]
   &  & Three-day & PM$_{2.5}$ trimmed at &  \\ 
   & Main analysis & temperature & 99th percentile & Tensor product \\[6pt]
  \hline
OR$_{10}$ & 1.05 (1.03, 1.06) & 1.03 (1.01, 1.05) & 1.04 (1.02, 1.06) & 1.05 (1.03, 1.10) \\ 
  OR$_{01}$ & 1.01 (0.99, 1.04) & 1.09 (1.07, 1.12) & 1.04 (1.02, 1.06) & 1.03 (0.98, 1.10) \\ 
  OR$_{11}$ & 1.09 (1.06, 1.12) & 1.14 (1.11, 1.18) & 1.12 (1.09, 1.14) & 1.10 (1.04, 1.19) \\ 
  RERI & 0.03 (0.01, 0.06) & 0.02 (-0.00, 0.05) & 0.03 (0.01, 0.05) & 0.02 (-0.03, 0.07) \\ 
   \hline
\end{tabular}
\caption{Odds ratios (OR) and relative excess risk due to interaction (RERI) results from the main analysis and three sensitivity analyses, as presented in Figure 3. Uncertainty intervals are 95\% Bayesian credible intervals. OR$_{10}$ indicates the independent effect of temperature, OR$_{01}$ indicates the independent effect of PM$_{2.5}$, and OR$_{11}$ indicates the combined effect of temperature and PM$_{2.5}$. All quantities displayed here are defined above in section 1.4.} 
\label{tab:or_results}
\end{table}


\newpage

\begin{thebibliography}{10}

\bibitem{ebi2021hot}
Ebi KL, Capon A, Berry P, Broderick C, de~Dear R, Havenith G, et~al.
\newblock Hot weather and heat extremes: health risks.
\newblock The lancet. 2021;398(10301):698-708.

\bibitem{weinberger2020estimating}
Weinberger KR, Harris D, Spangler KR, Zanobetti A, Wellenius GA.
\newblock Estimating the number of excess deaths attributable to heat in 297 United States counties.
\newblock Environmental Epidemiology. 2020;4(3):e096.

\bibitem{anderson2013heat}
Anderson GB, Dominici F, Wang Y, McCormack MC, Bell ML, Peng RD.
\newblock Heat-related emergency hospitalizations for respiratory diseases in the Medicare population.
\newblock American journal of respiratory and critical care medicine. 2013;187(10):1098-103.

\bibitem{fletcher2012association}
Fletcher BA, Lin S, Fitzgerald EF, Hwang SA.
\newblock Association of summer temperatures with hospital admissions for renal diseases in New York State: a case-crossover study.
\newblock American journal of epidemiology. 2012;175(9):907-16.

\bibitem{sun2018effects}
Sun Z, Chen C, Xu D, Li T.
\newblock Effects of ambient temperature on myocardial infarction: a systematic review and meta-analysis.
\newblock Environmental Pollution. 2018;241:1106-14.

\bibitem{layton2020heatwaves}
Layton JB, Li W, Yuan J, Gilman JP, Horton DB, Setoguchi S.
\newblock Heatwaves, medications, and heat-related hospitalization in older Medicare beneficiaries with chronic conditions.
\newblock PloS one. 2020;15(12):e0243665.

\bibitem{hasan2021effective}
Hasan F, Marsia S, Patel K, Agrawal P, Razzak JA.
\newblock Effective community-based interventions for the prevention and management of heat-related illnesses: a scoping review.
\newblock International journal of environmental research and public health. 2021;18(16):8362.

\bibitem{flynn2005older}
Flynn A, McGreevy C, Mulkerrin E.
\newblock Why do older patients die in a heatwave?
\newblock Qjm. 2005;98(3):227-9.

\bibitem{vaidyanathan2024heat}
Vaidyanathan A.
\newblock Heat-Related Emergency Department Visits—United States, May--September 2023.
\newblock MMWR Morbidity and Mortality Weekly Report. 2024;73.

\bibitem{khalaj2010health}
Khalaj B, Lloyd G, Sheppeard V, Dear K.
\newblock The health impacts of heat waves in five regions of New South Wales, Australia: a case-only analysis.
\newblock International archives of occupational and environmental health. 2010;83:833-42.

\bibitem{pachauri2014climate}
Pachauri RK, Allen MR, Barros VR, Broome J, Cramer W, Christ R, et~al.
\newblock Climate change 2014: synthesis report. Contribution of Working Groups I, II and III to the fifth assessment report of the Intergovernmental Panel on Climate Change.
\newblock IPCC; 2014.

\bibitem{stafoggia2008does}
Stafoggia M, Schwartz J, Forastiere F, Perucci C.
\newblock Does temperature modify the association between air pollution and mortality? A multicity case-crossover analysis in Italy.
\newblock American journal of epidemiology. 2008;167(12):1476-85.

\bibitem{li2017modification}
Li J, Woodward A, Hou XY, Zhu T, Zhang J, Brown H, et~al.
\newblock Modification of the effects of air pollutants on mortality by temperature: a systematic review and meta-analysis.
\newblock Science of the total environment. 2017;575:1556-70.

\bibitem{kioumourtzoglou2016pm2}
Kioumourtzoglou MA, Schwartz J, James P, Dominici F, Zanobetti A.
\newblock PM2.5 and mortality in 207 US cities: modification by temperature and city characteristics.
\newblock Epidemiology. 2016;27(2):221-7.

\bibitem{yitshak2018association}
Yitshak-Sade M, Bobb JF, Schwartz JD, Kloog I, Zanobetti A.
\newblock The association between short and long-term exposure to PM2.5 and temperature and hospital admissions in New England and the synergistic effect of the short-term exposures.
\newblock Science of the total environment. 2018;639:868-75.

\bibitem{tai2010correlations}
Tai AP, Mickley LJ, Jacob DJ.
\newblock Correlations between fine particulate matter (PM2. 5) and meteorological variables in the United States: Implications for the sensitivity of PM2. 5 to climate change.
\newblock Atmospheric environment. 2010;44(32):3976-84.

\bibitem{pryor2022physiological}
Pryor JT, Cowley LO, Simonds SE.
\newblock The physiological effects of air pollution: particulate matter, physiology and disease.
\newblock Frontiers in Public Health. 2022;10:882569.

\bibitem{basith2022impact}
Basith S, Manavalan B, Shin TH, Park CB, Lee WS, Kim J, et~al.
\newblock The impact of fine particulate matter 2.5 on the cardiovascular system: a review of the invisible killer.
\newblock Nanomaterials. 2022;12(15):2656.

\bibitem{basu2005temperature}
Basu R, Dominici F, Samet JM.
\newblock Temperature and mortality among the elderly in the United States: a comparison of epidemiologic methods.
\newblock Epidemiology. 2005;16(1):58-66.

\bibitem{maclure2000should}
Maclure M, Mittleman M.
\newblock Should we use a case-crossover design?
\newblock Annual review of public health. 2000;21(1):193-221.

\bibitem{kocher2013changes}
Kocher KE, Dimick JB, Nallamothu BK.
\newblock Changes in the source of unscheduled hospitalizations in the United States.
\newblock Medical care. 2013;51(8):689-98.

\bibitem{prism2024prism}
{PRISM Climate Group}. PRISM Climate Data; 2023.
\newblock Oregon State University. Accessed: 2024.
\newblock Available from: \url{http://prism.oregonstate.edu}.

\bibitem{daly2008physiographically}
Daly C, Halbleib M, Smith JI, Gibson WP, Doggett MK, Taylor GH, et~al.
\newblock Physiographically sensitive mapping of climatological temperature and precipitation across the conterminous United States.
\newblock International Journal of Climatology: a Journal of the Royal Meteorological Society. 2008;28(15):2031-64.

\bibitem{di2019ensemble}
Di Q, Amini H, Shi L, Kloog I, Silvern R, Kelly J, et~al.
\newblock An ensemble-based model of PM2.5 concentration across the contiguous United States with high spatiotemporal resolution.
\newblock Environment international. 2019;130:104909.

\bibitem{di2025pm25}
Di Q, Wei Y, Shtein A, Hultquist C, Xing X, Castro E, et~al.. {Daily, Monthly, and Annual PM2.5 Concentrations for the Contiguous United States, 1-km Grid (2000-2016)}. Harvard Dataverse; 2025.
\newblock Available from: \url{https://doi.org/10.7910/DVN/58C6HG}.

\bibitem{wang2018associations}
Wang Y, Zu Y, Huang L, Zhang H, Wang C, Hu J.
\newblock Associations between daily outpatient visits for respiratory diseases and ambient fine particulate matter and ozone levels in Shanghai, China.
\newblock Environmental pollution. 2018;240:754-63.

\bibitem{fernandez2018association}
Fern{\'a}ndez-Ni{\~n}o JA, Astudillo-Garc{\'\i}a CI, Rodr{\'\i}guez-Villamizar LA, Florez-Garcia VA.
\newblock Association between air pollution and suicide: a time series analysis in four Colombian cities.
\newblock Environmental health. 2018;17:1-8.

\bibitem{delaney2025extreme}
Delaney SW, Stegmuller A, Mork D, Mock L, Bell ML, Gill TM, et~al.
\newblock Extreme heat and hospitalization among older persons with Alzheimer disease and related dementias.
\newblock JAMA Internal Medicine. 2025;185(4):412-21.

\bibitem{gronlund2016vulnerability}
Gronlund CJ, Zanobetti A, Wellenius GA, Schwartz JD, O’Neill MS.
\newblock Vulnerability to renal, heat and respiratory hospitalizations during extreme heat among US elderly.
\newblock Climatic change. 2016;136(3):631-45.

\bibitem{di2017association}
Di Q, Dai L, Wang Y, Zanobetti A, Choirat C, Schwartz JD, et~al.
\newblock Association of short-term exposure to air pollution with mortality in older adults.
\newblock JAMA. 2017;318(24):2446-56.

\bibitem{nethery2023air}
Nethery RC, Josey K, Gandhi P, Kim JH, Visaria A, Bates B, et~al.
\newblock Air pollution and cardiovascular and thromboembolic events in older adults with high-risk conditions.
\newblock American journal of epidemiology. 2023;192(8):1358-70.

\bibitem{burkner2017r}
B{\"u}rkner PC.
\newblock An R package for Bayesian multilevel models using Stan.
\newblock Journal of Statistical Software. 2017;80(1):1-28.

\bibitem{rothman2012epidemiology}
Rothman KJ.
\newblock Epidemiology: an introduction.
\newblock Oxford university press; 2012.

\bibitem{knol2007estimating}
Knol MJ, van~der Tweel I, Grobbee DE, Numans ME, Geerlings MI.
\newblock Estimating interaction on an additive scale between continuous determinants in a logistic regression model.
\newblock International journal of epidemiology. 2007;36(5):1111-8.

\bibitem{vanderweele2014tutorial}
VanderWeele TJ, Knol MJ.
\newblock A tutorial on interaction.
\newblock Epidemiologic methods. 2014;3(1):33-72.

\bibitem{fayyad2024air}
Fayyad R, Josey K, Gandhi P, Rua M, Visaria A, Bates B, et~al.
\newblock Air pollution and serious bleeding events in high-risk older adults.
\newblock Environmental research. 2024;251:118628.

\bibitem{josey2023retrospective}
Josey K, Nethery R, Visaria A, Bates B, Gandhi P, Parthasarathi A, et~al.
\newblock Retrospective cohort study investigating synergism of air pollution and corticosteroid exposure in promoting cardiovascular and thromboembolic events in older adults.
\newblock BMJ open. 2023;13(9):e072810.

\bibitem{vandenbroucke2007strengthening}
Vandenbroucke JP, Elm Ev, Altman DG, G{\o}tzsche PC, Mulrow CD, Pocock SJ, et~al.
\newblock Strengthening the Reporting of Observational Studies in Epidemiology (STROBE): explanation and elaboration.
\newblock Annals of internal medicine. 2007;147(8):W-163.

\bibitem{r2021r}
{R Core Team}. R: A Language and Environment for Statistical Computing. Vienna, Austria; 2021.
\newblock Available from: \url{https://www.R-project.org/}.

\bibitem{areal2022effect}
Areal AT, Zhao Q, Wigmann C, Schneider A, Schikowski T.
\newblock The effect of air pollution when modified by temperature on respiratory health outcomes: A systematic review and meta-analysis.
\newblock Science of the Total Environment. 2022;811:152336.

\bibitem{anenberg2020synergistic}
Anenberg SC, Haines S, Wang E, Nassikas N, Kinney PL.
\newblock Synergistic health effects of air pollution, temperature, and pollen exposure: a systematic review of epidemiological evidence.
\newblock Environmental Health. 2020;19:1-19.

\bibitem{national2024weather}
{National Weather Service}. Weather Related Fatality and Injury Statistics; 2024.
\newblock Available from: \url{https://www.weather.gov/hazstat/}.

\bibitem{osborne2023trends}
Osborne TF, Veigulis ZP, Vaidyanathan A, Arreola DM, Schramm PJ.
\newblock Trends in heat related illness: nationwide observational cohort at the US Department of Veteran Affairs.
\newblock The Journal of Climate Change and Health. 2023;12:100256.

\bibitem{sorensen2022treatment}
Sorensen C, Hess J.
\newblock Treatment and prevention of heat-related illness.
\newblock New England Journal of Medicine. 2022;387(15):1404-13.

\bibitem{gauer2019heat}
Gauer R, Meyers BK.
\newblock Heat-related illnesses.
\newblock American family physician. 2019;99(8):482-9.

\bibitem{national2024heat}
{National Weather Service}. Heat Watch vs. Warning; 2024.
\newblock Available from: \url{https://www.weather.gov/safety/heat-ww}.

\bibitem{weinberger2018effectiveness}
Weinberger KR, Zanobetti A, Schwartz J, Wellenius GA.
\newblock Effectiveness of National Weather Service heat alerts in preventing mortality in 20 US cities.
\newblock Environment international. 2018;116:30-8.

\bibitem{toloo2013evaluating}
Toloo G, FitzGerald G, Aitken P, Verrall K, Tong S.
\newblock Evaluating the effectiveness of heat warning systems: systematic review of epidemiological evidence.
\newblock International journal of public health. 2013;58:667-81.

\bibitem{widerynski2017use}
Widerynski S, Schramm PJ, Conlon KC, Noe RS, Grossman E, Hawkins M, et~al.. Use of cooling centers to prevent heat-related illness: summary of evidence and strategies for implementation; 2017.

\bibitem{lin2013reducing}
Lin LY, Chuang HC, Liu IJ, Chen HW, Chuang KJ.
\newblock Reducing indoor air pollution by air conditioning is associated with improvements in cardiovascular health among the general population.
\newblock Science of the total environment. 2013;463:176-81.

\bibitem{taylor2020air}
Taylor-Clark TE.
\newblock Air pollution-induced autonomic modulation.
\newblock Physiology. 2020;35(6):363-74.

\bibitem{yamamoto2007evaluation}
Yamamoto S, Iwamoto M, Inoue M, Harada N.
\newblock Evaluation of the effect of heat exposure on the autonomic nervous system by heart rate variability and urinary catecholamines.
\newblock Journal of occupational health. 2007;49(3):199-204.

\bibitem{shah2019heart}
Shah AS, El~Ghormli L, Vajravelu ME, Bacha F, Farrell RM, Gidding SS, et~al.
\newblock Heart rate variability and cardiac autonomic dysfunction: prevalence, risk factors, and relationship to arterial stiffness in the treatment options for type 2 diabetes in adolescents and youth (TODAY) study.
\newblock Diabetes Care. 2019;42(11):2143-50.

\bibitem{florea2014autonomic}
Florea VG, Cohn JN.
\newblock The autonomic nervous system and heart failure.
\newblock Circulation research. 2014;114(11):1815-26.

\bibitem{jensen2015relationship}
Jensen MM, Brabrand M.
\newblock The relationship between body temperature, heart rate and respiratory rate in acute patients at admission to a medical care unit.
\newblock Scandinavian Journal of Trauma, Resuscitation and Emergency Medicine. 2015;23(Suppl 1):A12.

\bibitem{dvonch2009acute}
Dvonch JT, Kannan S, Schulz AJ, Keeler GJ, Mentz G, House J, et~al.
\newblock Acute effects of ambient particulate matter on blood pressure: differential effects across urban communities.
\newblock Hypertension. 2009;53(5):853-9.

\bibitem{sakurai2004plasma}
Sakurai M, Hamada K, Matsumoto K, Yanagisawa K, Kikuchi N, Morimoto T, et~al.
\newblock Plasma volume and blood viscosity during 4 h sitting in a dry environment: effect of prehydration.
\newblock Aviation, space, and environmental medicine. 2004;75(6):500-4.

\bibitem{sangkham2024update}
Sangkham S, Phairuang W, Sherchan SP, Pansakun N, Munkong N, Sarndhong K, et~al.
\newblock An update on adverse health effects from exposure to PM2.5.
\newblock Environmental Advances. 2024:100603.

\bibitem{hu2024effect}
Hu X, Fu H, Zhang L, Zhang Q, Xu T, Chen Y, et~al.
\newblock Effect of elevated temperatures on inflammatory cytokine release: an in vitro and population-based study.
\newblock Environment \& Health. 2024;2(10):721-8.

\bibitem{park2024mechanism}
Park M, Park S, Choi Y, Cho YL, Kim MJ, Park YJ, et~al.
\newblock The mechanism underlying correlation of particulate matter-induced ferroptosis with inflammasome activation and iron accumulation in macrophages.
\newblock Cell Death Discovery. 2024;10(1):144.

\bibitem{pope2016exposure}
Pope~III CA, Bhatnagar A, McCracken JP, Abplanalp W, Conklin DJ, O’Toole T.
\newblock Exposure to fine particulate air pollution is associated with endothelial injury and systemic inflammation.
\newblock Circulation research. 2016;119(11):1204-14.

\bibitem{briet2007endothelial}
Briet M, Collin C, Laurent S, Tan A, Azizi M, Agharazii M, et~al.
\newblock Endothelial function and chronic exposure to air pollution in normal male subjects.
\newblock Hypertension. 2007;50(5):970-6.

\bibitem{vybiral2005pyrogenic}
Vyb{\'\i}ral S, B{\'a}rczayov{\'a} L, Pe{\v{s}}anov{\'a} Z, Jansk{\`y} L.
\newblock Pyrogenic effects of cytokines (IL-1$\beta$, IL-6, TNF-$\alpha$) and their mode of action on thermoregulatory centers and functions.
\newblock Journal of Thermal Biology. 2005;30(1):19-28.

\bibitem{tran2023short}
Tran HM, Lin YC, Tsai FJ, Lee KY, Chang JH, Chung CL, et~al.
\newblock Short-term mediating effects of PM2.5 on climate-associated COPD severity.
\newblock Science of the Total Environment. 2023;903:166523.

\bibitem{ham2025mediation}
Ham D, Lim YH, Kim S, Kwon HJ, Bae S.
\newblock Mediation of Fine Particulate Matter on the Association Between Daily Temperature and Mortality.
\newblock Journal of Korean Medical Science. 2025;40(24).

\bibitem{zhu2019correlations}
Zhu J, Chen L, Liao H, Dang R.
\newblock Correlations between PM2. 5 and ozone over China and associated underlying reasons.
\newblock Atmosphere. 2019;10(7):352.

\bibitem{bell2007potential}
Bell ML, Kim JY, Dominici F.
\newblock Potential confounding of particulate matter on the short-term association between ozone and mortality in multisite time-series studies.
\newblock Environmental Health Perspectives. 2007;115(11):1591-5.

\bibitem{spangler2023does}
Spangler KR, Adams QH, Hu JK, Braun D, Weinberger KR, Dominici F, et~al.
\newblock Does choice of outdoor heat metric affect heat-related epidemiologic analyses in the US Medicare population?
\newblock Environmental Epidemiology. 2023;7(4):e261.

\bibitem{anderson2011heat}
Anderson GB, Bell ML.
\newblock Heat waves in the United States: mortality risk during heat waves and effect modification by heat wave characteristics in 43 US communities.
\newblock Environmental health perspectives. 2011;119(2):210-8.

\bibitem{hao2023national}
Hao H, Wang Y, Zhu Q, Zhang H, Rosenberg A, Schwartz J, et~al.
\newblock National cohort study of long-term exposure to PM2.5 components and mortality in Medicare American older adults.
\newblock Environmental Science \& Technology. 2023;57(17):6835-43.

\bibitem{masselot2022differential}
Masselot P, Sera F, Schneider R, Kan H, Lavigne {\'E}, Stafoggia M, et~al.
\newblock Differential mortality risks associated with PM2.5 components: a multi-country, multi-city study.
\newblock Epidemiology. 2022;33(2):167-75.

\bibitem{feng2024long}
Feng Y, Castro E, Wei Y, Jin T, Qiu X, Dominici F, et~al.
\newblock Long-term exposure to ambient PM2.5, particulate constituents and hospital admissions from non-respiratory infection.
\newblock Nature Communications. 2024;15(1):1518.

\bibitem{wood2013straightforward}
Wood SN, Scheipl F, Faraway JJ.
\newblock Straightforward intermediate rank tensor product smoothing in mixed models.
\newblock Stat Comput. 2013;23:341-60.

\bibitem{pedersen2019hierarchical}
Pedersen EJ, Miller DL, Simpson GL, Ross N.
\newblock Hierarchical generalized additive models in ecology: an introduction with mgcv.
\newblock PeerJ. 2019;7:e6876.

\end{thebibliography}
\end{document}